\documentclass[aps,prl,twocolumn,showpacs,floatfix]{revtex4}

\usepackage[english]{babel}
\usepackage{subfigure}
\usepackage{graphicx}

\begin{document}

\title{Suppression of decoherence via strong intra-environmental coupling}
\author{Luca Tessieri and Joshua Wilkie}
\affiliation{Department of Chemistry, Simon Fraser University, Burnaby,
             British Columbia V5A 1S6, Canada}

\date{9th July 2002}

\begin{abstract}
We examine the effects of intra-environmental coupling on decoherence by
constructing a low temperature spin--spin-bath model of an atomic impurity
in a Debye crystal. The impurity interacts with phonons of the crystal through
Jahn-Teller vibronic coupling. 
Anharmonic intra-environmental vibrational coupling is incorporated through
anti-ferromagnetic spin-spin interactions. The reduced density matrix of
the central spin representing the impurity is calculated by dynamically
integrating the full Schr\"{o}dinger equation for the spin--spin-bath model
for different thermally weighted eigenstates of the spin-bath. Exact
numerical results show that increasing the intra-environmental coupling
results in suppression of decoherence. This effect could play an important
role in the construction of solid state quantum devices such as quantum
computers.

\end{abstract}

\pacs{03.65.Yz, 03.67.Lx, 05.30.-d}

\maketitle

Intra-environmental coupling has customarily been neglected in theoretical
models of subsystem-environment interaction. The popular spin-boson
model~\cite{Leg,UW}, for example, assumes that the environment consists of
a set of non-interacting harmonic oscillators linearly coupled to a central
spin. The neglect of intra-environmental coupling is motivated more by
mathematical convenience than by physical insight. 

One failing of such models is that intra-environmental energy transfer can
only proceed by using the subsystem as an intermediate. In addition, it is
well known that the statistical properties of the energy eigenfunctions and
eigenspectra of strongly coupled (irregular) systems is qualitatively
different from that of uncoupled (regular) systems. The Wigner functions of
eigenvectors for irregular systems are almost uniform over the energetically
allowed classical phase space~\cite{Ber}. Those of regular systems are more
lumpy with energy localized in a subset of the available modes~\cite{Boh}.
Similarly, irregular eigenspectra show level repulsion while regular spectra
show level clustering~\cite{Ber2}. These spectral signatures have important
dynamical consequences~\cite{WB}. For these reasons coupled environments may
have decoherence properties which differ substantially from those predicted
by uncoupled oscillator models. 

Proposed new technologies such as quantum computing~\cite{Qcom}, laser control
of chemical reactions~\cite{Brum}, and molecular electronics~\cite{Ratn} all
require a qualitative understanding of the effects of decoherence and
dissipation for experimental implementation. Predictive theoretical studies
for such systems would also greatly benefit from dynamical methods which
accurately include the effects of intra-environmental coupling and
environmental memory effects. 

Unfortunately, exact theories such as the Feynman-Vernon influence functional
method~\cite{FV} and the Nakajima-Zwanzig master equation~\cite{Zwan} cannot
be easily applied. Approximate theories such as the Redfield~\cite{Opp},
completely-positive-dynamical-semigroup~\cite{dsg} and SRA~\cite{Wilk} master
equations need testing against exact results before they can be applied with
confidence. Thus, exact numerical calculations for subsystems interacting with
environments of a small number of degrees of freedom provide the only
practical method for exploring the effects of intra-environmental coupling.
Such studies would also allow us to test the accuracy of existing master
equations.

In this manuscript we report exact numerical results for the decoherence of
a central spin interacting with a spin-bath with intra-environmental coupling.
The model was constructed to represent an impurity in a thermal crystalline
solid; the impurity being the small quantum system and the solid the
environment. The first two electronic states of the impurity are vibronically
coupled to a number $n_s$ of coupled phonon modes of the crystal. At low
temperature the phonon modes can be represented as spin-1/2 modes~\cite{Sta}
(representing the first two states of an oscillator) with frequencies sampled
from a Debye spectrum with a frequency cutoff (set here at $\omega_D=1$).
With anharmonic phonon-phonon coupling effects included, our model Hamiltonian
then takes the form
\begin{equation}
\begin{array}{ccl}
H & = & \displaystyle
\frac{\omega_0}{2} \sigma_{z}^{(0)} + \beta \sigma_{x}^{(0)} +
\lambda_0 \sigma_{x}^{(0)} \sum_{j=1}^{n_s} \sigma_{x}^{(j)} \\
& + & \displaystyle
\sum_{j=1}^{n_s} \left[ \frac{\omega_j}{2} \sigma_{z}^{(j)} +
\beta \sigma_{x}^{(j)} \right] + \frac{\lambda}{2} \sum_{i\neq j=1}^{n_s}
\sigma_{x}^{(i)}\sigma_{x}^{(j)}
\end{array}
\label{HAM}
\end{equation}
where we arbitrarily chose $\omega_0=.8288$ as the frequency of the impurity,
$\beta=.01$ is the coefficient of a small anharmonic correction, and
$\lambda_{0}=1$ and $\lambda$ are the subsystem-environment and
intra-environmental coupling constants. Terms one, two and four of~(\ref{HAM})
represent the uncoupled modes of the subsystem (labeled with superscript 0)
and environment (labeled with superscripts 1 through $n_s$). The third term
in~(\ref{HAM}) couples the subsystem and environment, while the last term
couples the environment with itself. Here the sigmas represent Pauli matrices.
In our units $\hbar=1$. [Note that the same Hamiltonian could also represent
interacting nuclear or electronic spins in a solid.]

We calculated the reduced density matrix $\rho(t)$ of the impurity via the
formula
\begin{equation}
\rho(t) = \left( \begin{array}{cc}
                      \rho_{11}(t) & \rho_{10}(t)\\
                      \rho_{01}(t) & \rho_{00}(t)
                      \end{array} \right)
= \sum_{m=1}^{n_{eig}} p_{m} {\rm Tr}_{e}
\{|\psi_{m}(t)\rangle \langle \psi_{m}(t)|\}
\label{DENS}
\end{equation}
where 
\begin{displaymath}
p_{m} = \frac{\exp\{-\epsilon_{m}/kT\}}
{\sum_{m=1}^{n_{eig}} \exp\{-\epsilon_{m}/kT\}},
\end{displaymath}
$\epsilon_{m}$ and $|m\rangle_{e}$ are the energies and eigenvectors of
the isolated environment, and $kT$ is the temperature in units of energy.
The notation ${\rm Tr}_{e} \{ \hat{{\bf A}} \}$ indicates a trace of the
operator $\hat{{\bf A}}$ over the environmental degrees of freedom.
The states $|\psi_{m}(t)\rangle$ are evolved via the Schr\"{o}dinger
equation from initial states
\begin{equation}
|\psi_{m}(0)\rangle = |1\rangle_{0} \otimes |m \rangle_{e}
\label{istate}
\end{equation}
under Hamiltonian~(\ref{HAM}). The basis of eigenstates of the $\sigma_{z}$
operators was used to represent all states. The states $|0\rangle$ and
$|1\rangle$ represent down and up z-components of the spin, respectively.
Thus, the subsystem state $|1\rangle$ in Eq.~(\ref{istate}) means that the
impurity is initially in its excited state.

Equations~(\ref{DENS}) and~(\ref{istate}) represent an impurity in a thermal
solid which is excited by a fast laser pulse just prior to time $t=0$ (i.e.,
a sudden approximation) which then evolves while interacting with phonons in
the solid.

The calculations reported here are for $n_{s}=12$ bath spins. The ARPACK
linear algebra software~\cite{Arp} was used to calculate the lowest
$n_{eig}=20$ energies and eigenvectors of the isolated environment. A
temperature of $kT=.02$ was chosen such that no states with energy quantum
number $m$ higher than $n_{eig}$ would be populated at equilibrium.
The numerical solution of the Schr\"{o}dinger ordinary differential equation
for $|\psi_{m}(t)\rangle$ was calculated using an eighth order Runge-Kutta
routine~\cite{RK}.
Operations of the Hamiltonian~(\ref{HAM}) on the wavevector were calculated
via repeated application of Pauli matrix multiplication routines. For example 
\begin{displaymath}
\langle j_{n_s},\ldots,j_i,\ldots,j_1| \sigma_{x}^{(i)} |\psi \rangle =
\langle j_{n_s},\ldots,\bar{j_i},\ldots,j_1| \psi \rangle
\end{displaymath}
for all sets of $j_l=0,1$, $l=1,\ldots ,n_s$ and where $\bar{j_i}=1$ if
$j_i=0$ and $\bar{j_i}=0$ if $j_i=1$. Thus, an operation of $\sigma_{x}^{(i)}$
simply rearranges the components of $|\psi\rangle$. States of the basis 
can be represented by integers $j = j_{1} + j_{2}2 +\ldots + j_{i}2^{i-1} +
\ldots + j_{n_s}2^{n_s-1}$ and since integers are represented in binary form
on a computer, the mapping $j \rightarrow j' = j_{1} + j_{2}2 + \ldots +
\bar{j_i}2^{i-1} + \ldots + j_{n_s}2^{n_s-1}$ under $\sigma_{x}^{(i)}$ can be
calculated very simply using Fortran binary-operation intrinsic functions.
Operations for $\sigma_{y}^{(i)}$ and  $\sigma_{z}^{(i)}$ are also
straightforward.

\begin{figure}
\caption{Subsystem entropy $S$ versus time $t$ plotted for different values
of $\lambda$}
\includegraphics[width=3.4in]{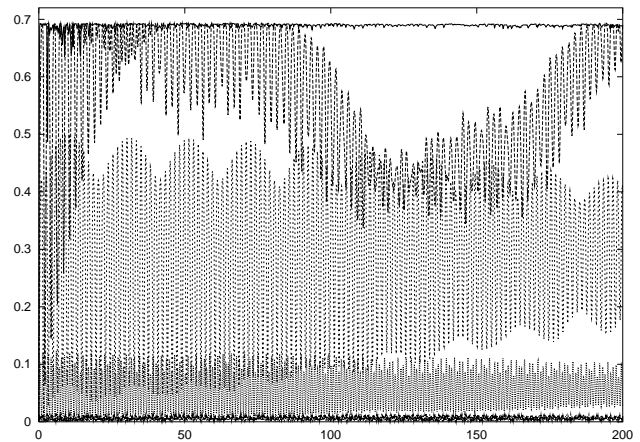}
\label{fig1}
\end{figure}

In Fig.~\ref{fig1} we show the calculated subsystem entropy
\begin{equation}
\begin{array}{ccl}
S(t) & = & \displaystyle -{\rm Tr}_s\{\rho(t)\ln \rho(t)\}
 =  -\frac{1}{2} \ln \{ \det [\rho(t)] \} \\
 & - & \displaystyle \frac{1}{2}
\sqrt{1 - 4 \det [\rho(t)]} \ln \frac{1 + \sqrt{1 - 4 \det [\rho(t)]}}
{1 - \sqrt{1 - 4 \det [\rho(t)]}}
\end{array}
\label{S}
\end{equation}
where $\det [\rho(t)] = \rho_{11}(t) \rho_{00}(t) - \rho_{10}(t)\rho_{01}(t)$,
for various values of the intra-environmental coupling $\lambda$.
For $\lambda=0$ (solid curve) the entropy rapidly approaches its maximum value
of $\ln(2) \simeq 0.693147 \ldots$ which corresponds to the diagonalised
reduced density matrix
\begin{eqnarray*}
\rho_{00}(t) = \rho_{11}(t) = \frac{1}{2} \\
\rho_{10}(t) = \rho_{01}(t) = 0 .
\end{eqnarray*}
As $\lambda$ increases through 1 (long dashed curve), 2 (short dashed curve),
4 (dotted curve), and 8 (dot-dashed curve), the entropy approaches a
smaller asymptotic value. This strongly suggests that increased
intra-environmental coupling suppresses decoherence.

To confirm this we compare the dynamics of the three subsystem spin components
\begin{eqnarray*}
X(t) & = & {\rm Tr}_{s} \{\sigma_{x}^{(0)} \rho(t)\}
= \rho_{10}(t) + \rho_{01}(t)\\
Y(t) & = & {\rm Tr}_{s} \{\sigma_{y}^{(0)} \rho(t)\}
= i (\rho_{10}(t) - \rho_{01}(t))\\
Z(t) & = & {\rm Tr}_{s} \{\sigma_{z}^{(0)} \rho(t)\}
= \rho_{11}(t) - \rho_{00}(t)
\end{eqnarray*}
calculated for Hamiltonian~(\ref{HAM}) with those of the impurity evolving
in isolation (i.e., $\lambda_{0}=0$).
\begin{figure}
\subfigure[$X(t)$]{
\label{fig2a}
\includegraphics[width=3.4in]{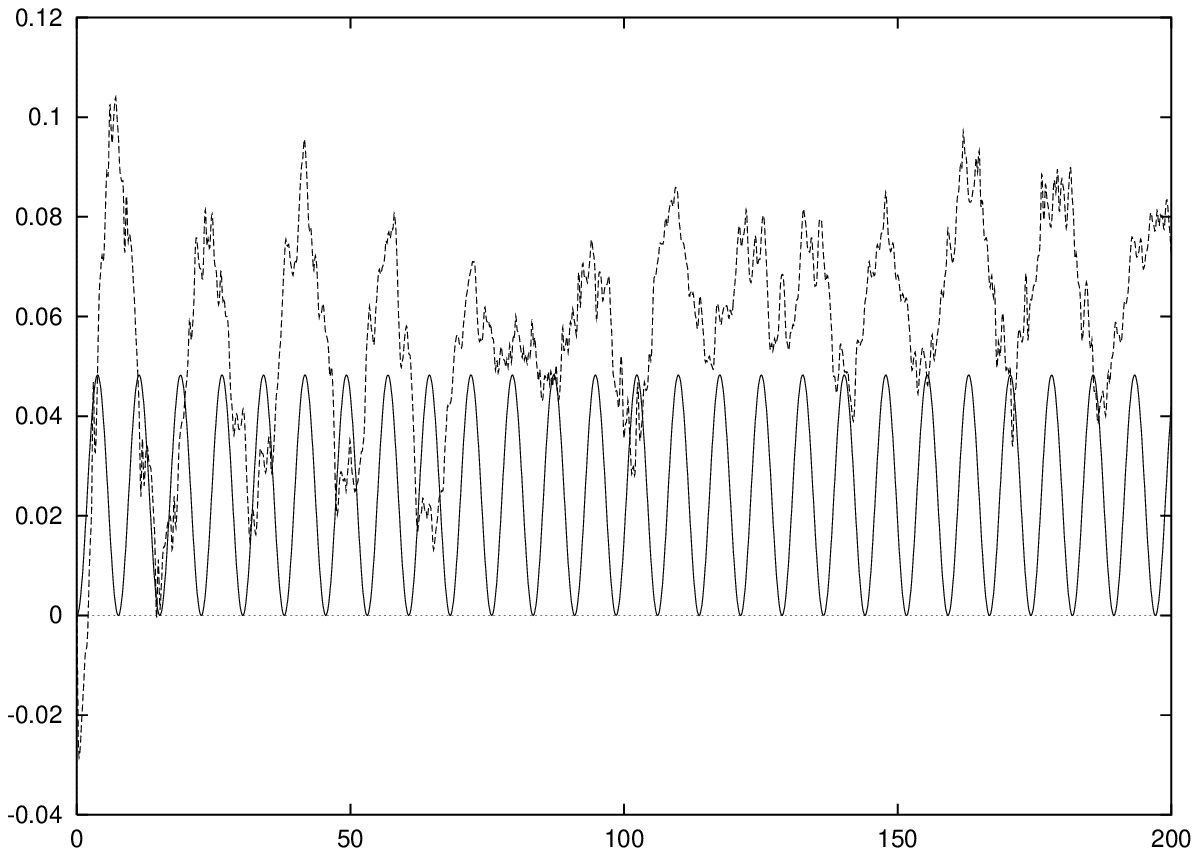}}
\subfigure[$Y(t)$]{
\label{fig2b}
\includegraphics[width=3.4in]{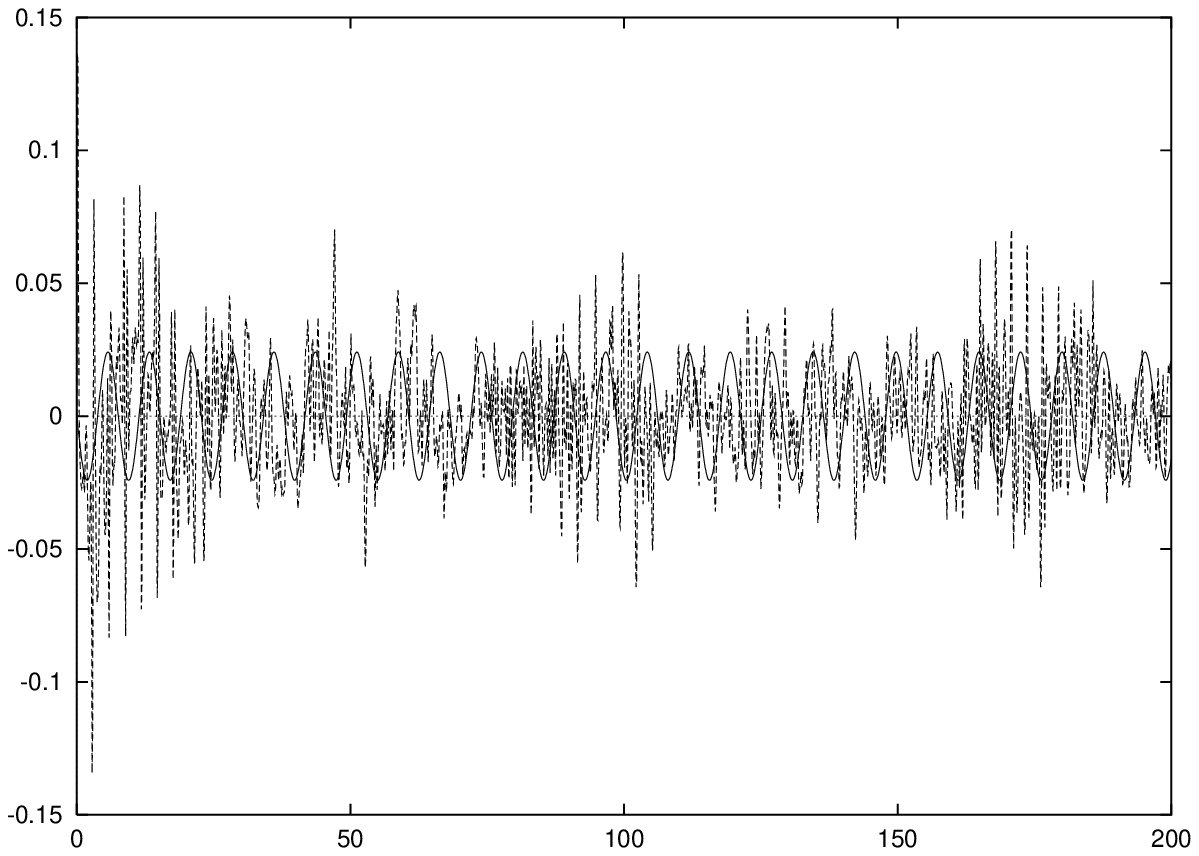}}
\subfigure[$Z(t)$]{
\label{fig2c}
\includegraphics[width=3.4in]{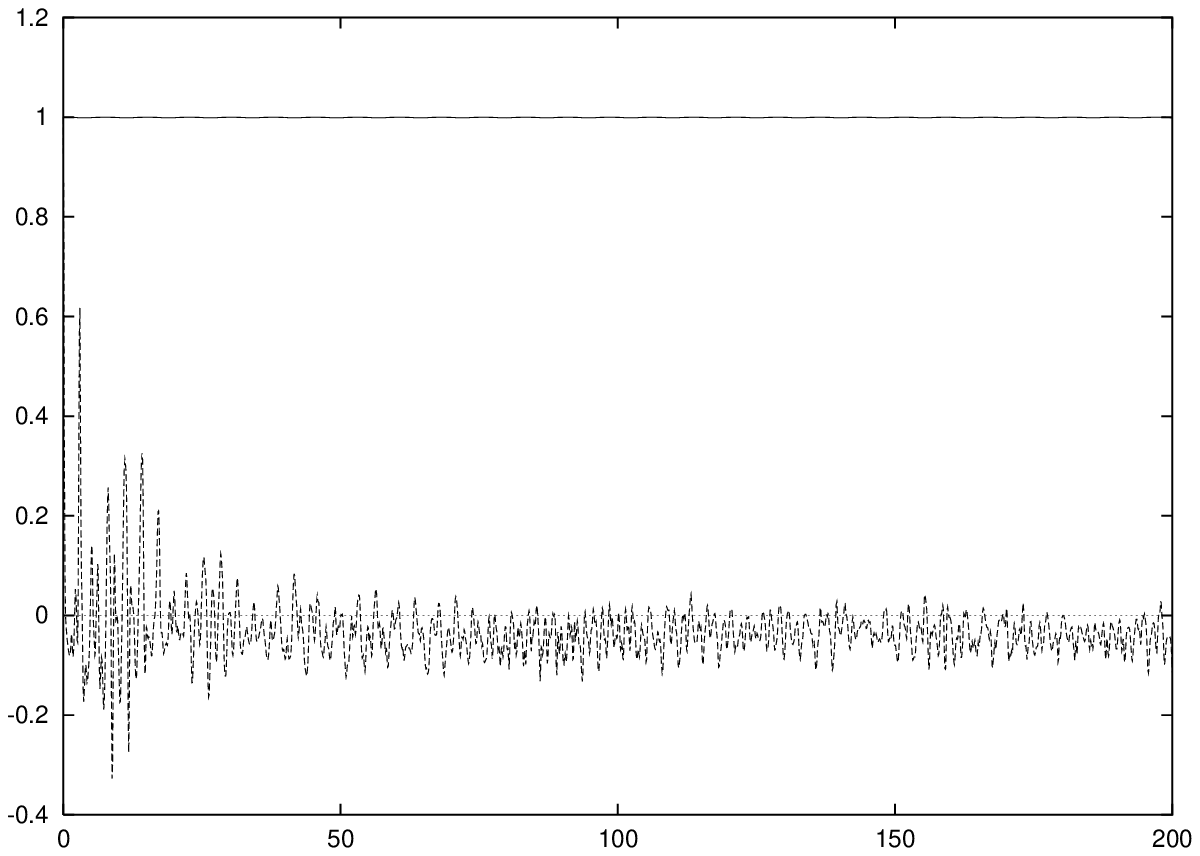}}
\caption{$x$, $y$, and $z$ component of the spin versus $t$ for
         $\lambda = 0$}
\label{fig2}
\end{figure}
\begin{figure}
\subfigure[$X(t)$]{
\label{fig3a}
\includegraphics[width=3.4in]{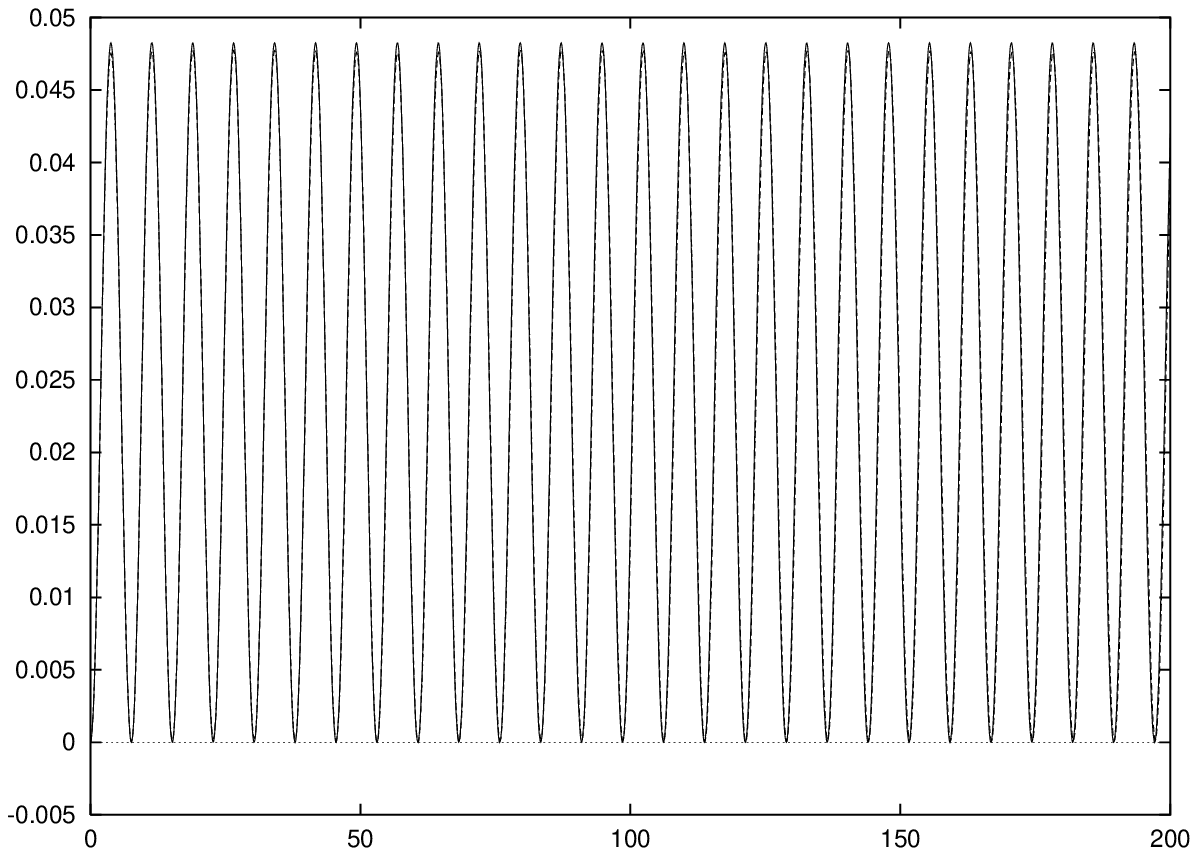}}
\subfigure[$Y(t)$]{
\label{fig3b}
\includegraphics[width=3.4in]{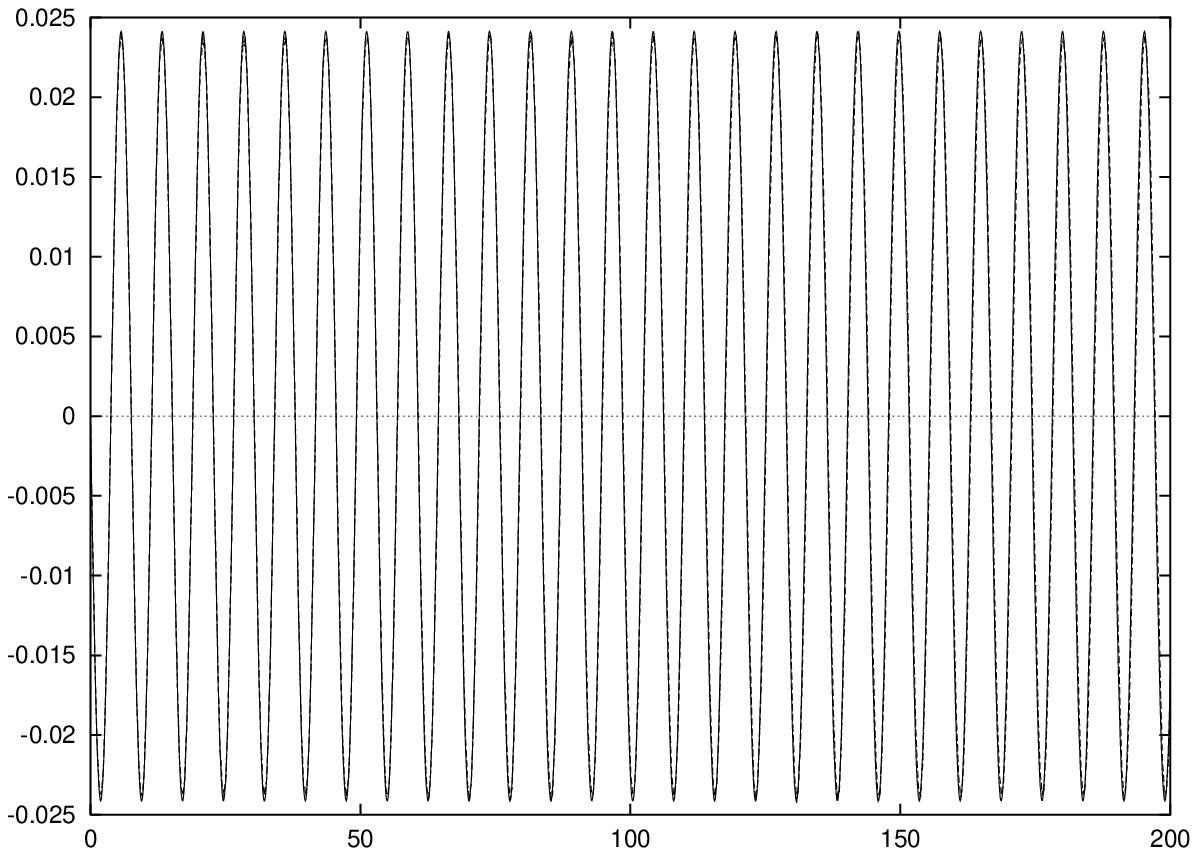}}
\subfigure[$Z(t)$]{
\label{fig3c}
\includegraphics[width=3.4in]{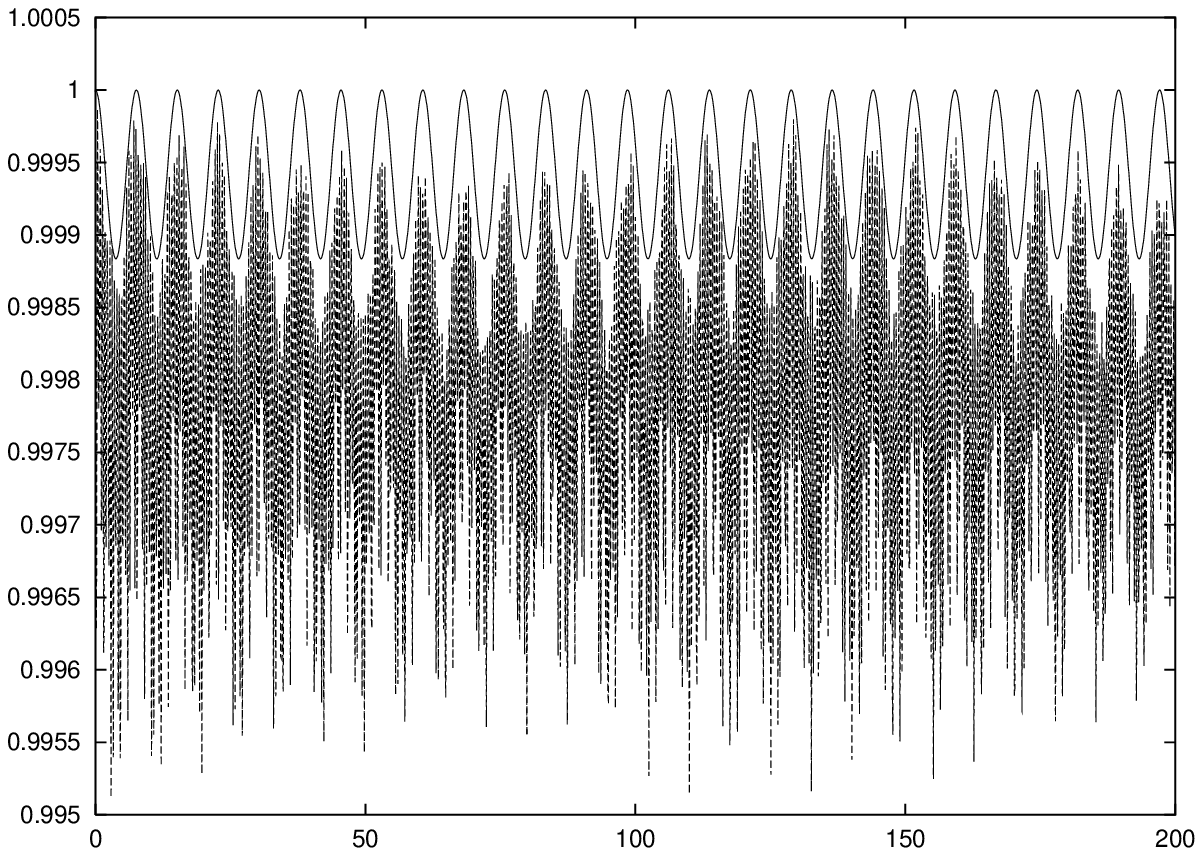}}
\caption{$x$, $y$, and $z$ component of the spin versus $t$ for
         $\lambda = 8$}
\label{fig3}
\end{figure}
In Fig.~\ref{fig2} we show the time evolution of the central spin in
the absence of interaction with the bath, i.e., for $\lambda_{0}=0$ (solid
line) and the evolution of the same spin for $\lambda_{0}=1$ and
$\lambda=0$ (dashed line), i.e., when the central spin is coupled to a
bath of non-interacting spins.
As can be seen from Fig.~\ref{fig2c}, the system undergoes rapid
decoherence when the interaction with the bath is turned on.
In Fig.~\ref{fig3} we show again the behaviour of $X(t)$, $Y(t)$, and
$Z(t)$ for $\lambda_{0}=0$ (solid line, isolated spin) and for $\lambda_{0}=1$
and $\lambda = 8$ (dashed line, central spin coupled to a bath of strongly
interacting modes).
Here the agreement between the coupled and isolated dynamics is much
better than in the previous case; in fact, the only significant discrepancy
appears in the behaviour of $Z(t)$, whereas the plots of $X(t)$ and $Y(t)$
for $\lambda_{0} = 0$ and $\lambda_{0} = 1$ are almost indistinguishable.

Thus, at least in this simple model, increasing anharmonic intra-environmental
coupling incrementally suppresses decoherence. There is a relatively simple
explanation for the observed behaviour. Define an environmental super-spin
with components
\begin{eqnarray*}
\Sigma_{x} = \sum_{j=1}^{n_s} \sigma_{x}^{(j)}, &
\Sigma_{y} = \displaystyle \sum_{j=1}^{n_s} \sigma_{y}^{(j)}, &
\Sigma_{z} = \sum_{j=1}^{n_s} \sigma_{z}^{(j)},
\end{eqnarray*}
in terms of which Hamiltonian~(\ref{HAM}) can be rewritten as
\begin{equation}
\begin{array}{ccl}
H & = & \displaystyle
\frac{\omega_{0}}{2} \sigma_{z}^{(0)} + \beta \sigma_{x}^{(0)}
+ \lambda_{0} \sigma_{x}^{(0)} \Sigma_{x} \\
& + & \displaystyle \sum_{j=1}^{n_{s}} \frac{\nu_{j}}{2}
\sigma_{z}^{(j)} + \frac{\Omega}{2} \Sigma_{z} 
+ \beta \Sigma_{x} + \frac{\lambda}{2}[\Sigma_{x}^{2}-n_{s}{\bf 1}]
,\label{HAM2}\end{array}
\end{equation}
where $\Omega = \sum_{j=1}^{n_{s}} \omega_{j}/n_{s}$ is the average
environment frequency and $\nu_{j} = \omega_{j}-\Omega$.
When $\lambda \gg \Omega$, the last term in~(\ref{HAM2}) dominates the
bath Hamiltonian. This entails that the bath eigenstates are nearly
eigenvectors of $\Sigma_{x}$; specifically, the bath eigenstates with
lowest energy correspond to the lowest eigenvalues of $\Sigma_{x}$.
Since the initial conditions~(\ref{istate}) are low-energy bath eigenstates,
the evolving dynamical states $|\psi_{m}(t)\rangle$ remain close to
eigenstates of $\Sigma_{x}$ with small eigenvalues; as a consequence
the central spin becomes nearly decoupled from the environment and its
evolution is determined by the effective Hamiltonian
\begin{displaymath}
H_{s} \simeq  (\omega_{0}/2) \sigma_{z}^{(0)}
+ ( \beta + \lambda_{0} s_{x} ) \sigma_{x}^{(0)}
\end{displaymath}
where $s_{x}$ is a (typically small) eigenvalue of $\Sigma_{x}$.

Our study strongly suggests that intra-environmental coupling has as important
an effect on decoherence as temperature or subsystem-environmental coupling.
Environments of $N$ (coupled oscillator) phonons should have similar
decoherence properties. Since the Wigner functions of the energy eigenstates
of strongly coupled systems are nearly uniform over the classical energy
surface~\cite{Ber}, energy is distributed equally among all modes. Thus
displacements from equilibrium of any phonon mode must be small at low
temperature, especially in the thermodynamic limit. Since the coupling of an
impurity to a phonon is through its displacement coordinate, this coupling
will also be small. By contrast, energy distribution in uncoupled oscillator
systems is non-uniform and may be localized in a small number of
modes~\cite{Boh}. Hence vibronic coupling to these modes will be strong as
will decoherence.

We have obtained similar results for both larger and smaller numbers of bath
spins and also in the case of ferromagnetic intra-environmental coupling.
We believe that our model is a reasonably accurate representation of an
impurity in a low-temperature crystal. If such strong intra-environmental
coupling exists in nature, it could be exploited as a platform for
quantum computing. We are currently developing methods for the study of
higher-temperature systems where environmental modes are modeled by coupled
harmonic oscillators.

The authors gratefully acknowledge the financial support of the Natural
Sciences and Engineering Research Council of Canada.

\end{document}